\documentclass{svjour3}                     
\smartqed  
\usepackage{graphicx}
\usepackage{psfrag}
\usepackage{epsfig}
\usepackage{amsmath}
\usepackage{color}
\journalname{Experiments in Fluids}
\begin{document}

\title{Dead time effects in laser Doppler anemometry measurements}

\author{Clara M. Velte         \and
        Preben Buchhave     \and
        William K. George
}

\institute{Clara M. Velte \at
            Department of Mechanical Engineering\\
            Technical University of Denmark\\
              Nils Koppels All\'{e} Bldg. 403 \\
              2800 Kgs. Lyngby, Denmark\\
              Tel.: +45-45254342\\
              Fax: +45-45930663\\
              \email{clara@alumni.chalmers.se}
              \and
            Preben Buchhave \at
            Intarsia Optics\\
            S{\o}nderskovvej 3\\
            3460 Birker{\o}d, Denmark
           \and
           William K. George \at
           Department of Mechanical and Aerospace Engineering\\
           Princeton University\\
           Princeton, NJ 08544\\
            and\\
            Department of Aeronautics\\
            Imperial College London\\
            South Kensington Campus\\
            London SW7 2AZ\\
}

\date{Received: date / Accepted: date}

\maketitle

\begin{abstract}
We present velocity power spectra computed by the so-called direct method from burst type laser Doppler anemometer (LDA) data, both measured in a turbulent round jet and generated in a computer. Using today's powerful computers we have been able to study more properties of the computed spectra than was previously possible, and we noted some unexpected features of the spectra that we now attribute to the unavoidable influence of a finite measurement volume (MV). The most prominent effect, which initially triggered these studies, was the appearance of damped oscillations in the higher frequency range, starting around the cut-off frequency due to the size of the MV. Using computer generated data mimicking the LDA data, these effects have previously been shown to appear due to the effect of dead time, i.e., the finite time during which the system is not able to acquire new measurements. These dead times can be traced back to the fact that the burst-mode LDA cannot measure more than one signal burst at a time. Since the dead time is approximately equal to the residence time for a particle traversing a measurement volume, we are dealing with widely varying dead times, which, however, are assumed to be measured for each data point. In addition, the detector and processor used in the current study introduce a certain amount of fixed processing and data transfer times, which further contribute to the distortion of the computed spectrum. However, we show excellent agreement between a measured spectrum and our modeled LDA data thereby confirming the validity of our model for the LDA burst processor.

\end{abstract}

\section{Introduction}
\label{intro}

Computation of power spectra of turbulent flows measured with the laser Doppler anemometer (LDA) has historically proven to be a difficult task. The source of the LDA data is light scattered from randomly dispersed particles carried by the fluid through a measurement volume (MV) formed by the intersection of two laser beams. The random, but velocity dependent, data acquisition constitutes one of the main signal processing challenges. Although a number of methods have been developed to process the resulting randomly sampled velocity data, see e.g. the structured review in~\cite{2Albrecht} and the testing of algorithms and discussions in~\cite{10}, no method has come out a clear winner when it comes to computation of velocity power spectra. Each method has its own problems, and the resulting spectra are subject to distortions due to a number of effects inherent in real-life LDA instruments.

In this paper we focus our investigation on the effect of a finite measurement volume on the power spectrum measured by a burst-type LDA and computed by the so-called direct method as described first by Gaster and Roberts in 1977~\cite{GR14}. It is well known that the finite size of the measurement volume leads to limitations in the spatial and temporal resolution, but the specific effects on the computed power spectra has so far not been investigated in detail. Dead time effects were suspected to play a role already in the 1960s when Gaster and Roberts were processing randomly sampled data while studying results produced by the `Particle Velocity Meter' developed at the National Physical Laboratory, and when their work~\cite{14} was applied to the Malvern photon correlator, but the limited computing power available at that time prevented further analysis~\cite{14b}. The dead time problem has also been treated for X-ray photon counters, where white power spectra were expected, but measured spectra displayed oscillatory behavior due to a constant dead time imposed by the photo detector~\cite{6,7}.

Recently, we have analyzed in more detail the effect of a fixed dead time on the shape of a computed power spectrum~\cite{56}. We investigated the dead time effects in the general case of randomly sampled data, where the sampling and sampled processes could be assumed statistically independent. Filtering and fixed dead time effects were deduced analytically and tested using computer simulations. The effect of a fixed detector dead time on a von K\'{a}rm\'{a}n spectrum was confirmed by comparing our analytical model to computer simulations with dead time included.

The current work is focused on the special case of the burst-type LDA processor and builds upon the previous work~\cite{56} by including a number of additional effects inherent in LDA measurements. Naturally, the LDA has a more complex behavior when measurements are performed in a turbulent flow where the velocities vary substantially. The dead time is associated with the time a particle is present in the measurement volume and the processor is busy acquiring the velocity information. This essentially precludes a measurement within the residence time and results in a distortion of the distribution of short time lags between measurements. The dead time model presented in the current work has therefore been extended to include variable dead times. This variable dead time is provided by correctly working LDA burst processors in the form of residence times, which will also be able to account for effects due to particle interference in the measuring volume. The exact shape of the spectrum distortion is in fact very sensitive to the type of dead time, i.e., whether we assume a dead time of a fixed or widely varying length. This high sensitivity is also reflected in whether we assume a dead time that is affected by samples arriving within an ongoing dead time. If incoming samples within the dead time are simply ignored the detector is non-paralyzable. If they instead add to the current dead time, the detector can for high data rates become paralyzed, i.e., the detector is paralyzable. We argue in the paper that the burst-type LDA is a special example of a paralyzable dead time caused by interference of two or more particles being present simultaneously in the measuring volume.

In this paper we develop a model for the type of dead time to be expected from a burst-type LDA, and we apply this model to computer generated velocity data derived from a model von K\'{a}rm\'{a}n turbulence power spectrum. We then compute the resulting power spectrum and compare this spectrum to a spectrum computed from real LDA measurements in a free turbulent jet. In all computations in this paper we apply the so-called direct residence time weighted (RTW) algorithms that essentially remove the correlation between sample rate and velocity magnitude. We have previously shown that the only \textit{in principle} correct way to process the signals from a burst-type LDA is to multiply each velocity data point by the time it is actually present in the measurement volume, the residence time~\cite{10,12,11,4}. The result is exactly equivalent to a time average of a conventional regularly sampled signal, and the resulting statistical moments are unbiased by the correlation existing between the sample rate and the instantaneous velocity magnitude. Furthermore, unlike slotting or interpolation methods that require a large number of slots to avoid aliasing and averaging effects, the direct method is a simple estimator and it therefore brings out the essential information for studying dead time phenomena. We want to emphasize already here that our results apply specifically to the direct method and that other dead time effects are likely to appear in other methods used for computing the power spectrum.

By realistic assumptions about the size of the measuring volume and the noise generated in the photo detection process, we achieve a near perfect match between the computer generated power spectrum and the spectrum from the free jet showing that our assumptions regarding the operation of the burst-type processor are realistic and accurate. In the following, we first describe the way the data is processed and explain the function of the burst-mode LDA processor. We continue in Section~\ref{sec:dtmodels} by describing how the special case of the LDA differs from the previous work~\cite{56}, introduce computer generated data used to investigate the properties of the power spectrum and develop the model for the dead time in the LDA. To validate our model, we then describe the experimental LDA setup and the properties of the measured power spectrum, emphasizing the motivation for the work. Finally, in Section~\ref{sec:commeas} we compare the measured spectrum and a computer generated spectrum showing that the model convincingly explains the properties of the measured spectrum.

\section{Processing of LDA data}

Data sampling by laser Doppler anemometers display added complexity compared to equidistantly sampled data due to the fact that the sampling process and the measured fluctuating velocity component are correlated; the sample rate is in principle proportional to the magnitude of the velocity vector, the so-called velocity bias. This leads to two concerns of interest here:

\begin{enumerate}
\item The statistical moments should be computed using residence time correction. This residence time correction will be applied to all calculations in the following. The residence time weighted spectral estimator used is given by~\cite{10,12}:
    \begin{equation}\label{eqn:specest}
    S_{0,rt}(f) = \frac{T}{\left (\sum_{k=0}^{N-1}\Delta t_{s,k} \right )^2} \left | \sum_{k=0}^{N-1}u(t_k)\Delta t_{s,k} e^{-i2\pi f t_k} \right |^2 \end{equation}
    Note that this estimator includes self-product that cause a constant spectral offset but that this offset may be subtracted.
\item The velocity measurement is the result of the processing of the digitized Doppler signal, sampled while a particle is in the measurement volume. Thus, LDA is an example of a burst type measurement displaying both top-hat averaging during the sampling time $\Delta t_p$ and a dead time $\Delta t_d$ resulting because a measurement can only be transmitted to the subsequent data processor after the signal has been sampled and processed. The LDA measurements display an added complication compared to the theory developed in~\cite{56} in that the dead time will, due to the velocity fluctuations and the varying path lengths for particles traversing the measurement volume, suffer large variations from sample to sample. Thus, we have a case of a, usually wide, distribution of dead times.
\end{enumerate}

\section{The burst processor} \label{sec:2}

In the previous work on the effect of fixed dead times on randomly sampled power spectral estimates~\cite{56}, the signal was assumed to originate from a Doppler modulated electronic pulse similar to the manner in which many LDA burst processors function.

\begin{figure*}[!h]
  \center{\includegraphics[width=0.5\textwidth]{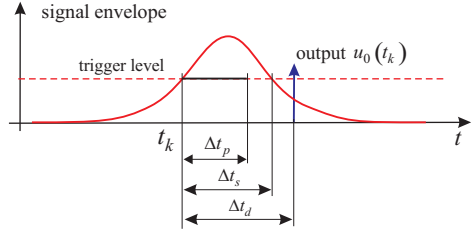}}
\caption{The sampling process (from~\cite{56}).} \label{fig:Fig1}
\end{figure*}

As illustrated in Figure~\ref{fig:Fig1}, a signal is detected by a burst detector during the time $\Delta t_s$ the signal level exceeds the trigger level, commonly referred to as the residence time or transit time. During the burst, the signal is digitized and processed to provide one numerical velocity output for each burst. The processing of the signal is assumed to begin immediately after the burst detection and last for a processing time $\Delta t_p$, which may be shorter than $\Delta t_s$ depending on the processor. One may also include time for transfer of the processed data to a subsequent data processor. During this transfer time the system is assumed unable to acquire new data. The total time from initial burst detection to the instant the processor is again ready for measurement is denoted the dead time, $\Delta t_d$.

The detector and signal processor were assumed to cause the signal to be averaged during the processing time $\Delta t_p$. This is mathematically represented by the true velocity signal $u(t)$ being convolved with a top hat function of width $\Delta t_p$. The Fourier transform of the filtered signal is thus
\begin{equation}\label{eqn:one}
\tilde{u}_{\Delta t_p}(f) = \tilde{u}(f)\cdot \mathrm{sinc} (\pi f \Delta t_p)
\end{equation}
resulting in a power spectrum
\begin{equation}\label{eqn:two}
S_{0,\Delta t_p}(f) = \frac{\overline{u^2_{\Delta t_p}}}{\nu} + S(f)\cdot \mathrm{sinc}^2 (\pi f \Delta t_p)
\end{equation}
with a sinc$^2$ transmission function due to the filtering. The first term is a constant offset term that does not include any spectral information and may be ignored.

\begin{figure*}[!h]
  \center{\includegraphics[width=0.5\textwidth]{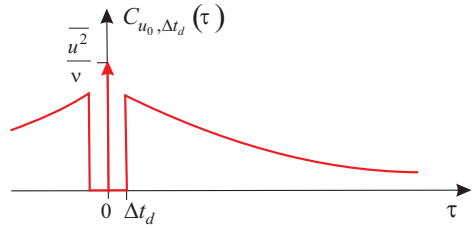}}
\caption{Autocovariance function with dead time effect (from~\cite{56}).} \label{fig:Fig7}
\end{figure*}

In~\cite{56} dead time was introduced in correlation space by multiplying the autocovariance function (ACF) by an inverted top hat function of width $2 \Delta t_d$ around the origin, see Figure~\ref{fig:Fig7}. Neglecting the effect of averaging of the signal during the processing time $\Delta t_p$, which is generally small in comparison, the obtained power spectrum could be expressed as
\begin{equation}\label{eqn:specdead}
S_{0,\Delta t_d}(f) = \frac{\overline{u^2_{\Delta t_d}}}{\nu_0} + S(f)\otimes [ \delta (f) - 2\Delta t_d \mathrm{sinc} (2\pi f \Delta t_d)]
\end{equation}
The first term is again a constant offset term that does not include any spectral information and $\nu_0$ is the reduced sample rate due to dead time, $\nu_0 = \nu e^{-\nu \Delta t_d}$~\cite{56}.

\section{Dead time model}\label{sec:dtmodels}

\subsection{Paralyzable vs. non-paralyzable detectors}\label{sec:paral}

\begin{figure*}[!h]
  \center{\includegraphics[width=0.5\textwidth]{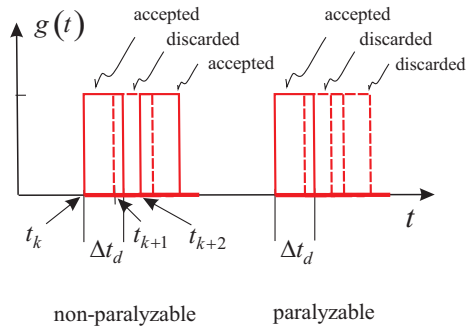}}
\caption{Illustration of non-paralyzable and paralyzable dead time.} \label{fig:Fig10}
\end{figure*}

The detailed functioning of detector and signal processor can have a significant impact on the description of the dead time and on the measured power spectrum. In~\cite{56} we describe an example of a so-called non-paralyzable fixed dead time where the system is insensitive to all samples arriving within the dead time. Samples that would be measured by an ideal detector are simply ignored if they arrive within the dead time. As soon as the system has finished the measurement and has passed the result to the data processor it is ready for a new measurement. However, many measurement systems behave in a different way, as a so-called paralyzable detector. In these cases, a sample arriving within a dead time is registered by the detector. The detector senses the new sample, and the dead time is increased by the dead time of the new sample. Thus, if the sample rate is high enough, the samples can in principle arrive so closely spaced that the system cannot recover, and the dead time continues to grow - the system is paralyzed. The sketch in Figure~\ref{fig:Fig10} illustrates the two cases.

\subsection{Computer generated data with dead time}\label{sec:CGdata}

To illustrate the effect on the power spectrum of these two dead time models, we have used a model von K\'{a}rm\'{a}n spectrum with an exponential tail, with properties matching the measured hot-wire turbulence power spectrum discussed more in detail in Section~\ref{sec:measuredspectrum}, Figure~\ref{fig:Fig12},
\begin{equation}\label{eqn:vkspectrum}
S_{vK}(f) = \frac{1}{62.5}\cdot \frac{1}{\left ( 1 + (f/45)^2 \right )^{5/6}} \exp \left ( - (f/2500)^{4/3} \right )
\end{equation}
to generate random data in a computer and provide them with the characteristics of an LDA signal subject to dead time effects. Briefly outlined the method is the following: First a series of evenly distributed random frequency values are passed through a frequency filter with the von K\'{a}rm\'{a}n response. The filtered frequencies are then converted into a time series, $u_{primary}$, by an FFT process. This primary velocity time series is then used as input to a Poisson process to provide a series of arrival times
\begin{equation}\nonumber
t_a = \mathrm{Poisson} \left \{ |u_{primary}(t)|,\mu \right \}
\end{equation}
where $\mu$ is an adjustable parameter ensuring that the Poisson process provides primarily zero or one. The sample rate will then be proportional to the magnitude of the velocity.

The arrival times are then used to provide the randomly sampled velocity data
\begin{equation}\nonumber
u(t_a) = u_{primary}(t_a)
\end{equation}
and the corresponding residence times
\begin{equation}\nonumber
\Delta t_s(t_a) = d_{MV}/|u_{primary}(t_a)|
\end{equation}
where $d_{MV}$ is the diameter of the measurement volume. We have tested varying the MV size corresponding to likely variations in particle path through the measuring volume, with the result of a mere constant of  $3\,dB$ added to the noise floor. The generated arrival times have a resolution given by the primary time series. We then pass these data through a dead time filter which simply omits any realization that occurs within the prescribed dead time of a previous realization. Finally, we process these computer generated data through the same spectral estimator as we used for the measured velocity data, Eq.~(\ref{eqn:specest}).

\begin{figure*}[!h]
\begin{minipage}{0.5\linewidth}
  \center{\includegraphics[width=1.0\textwidth]{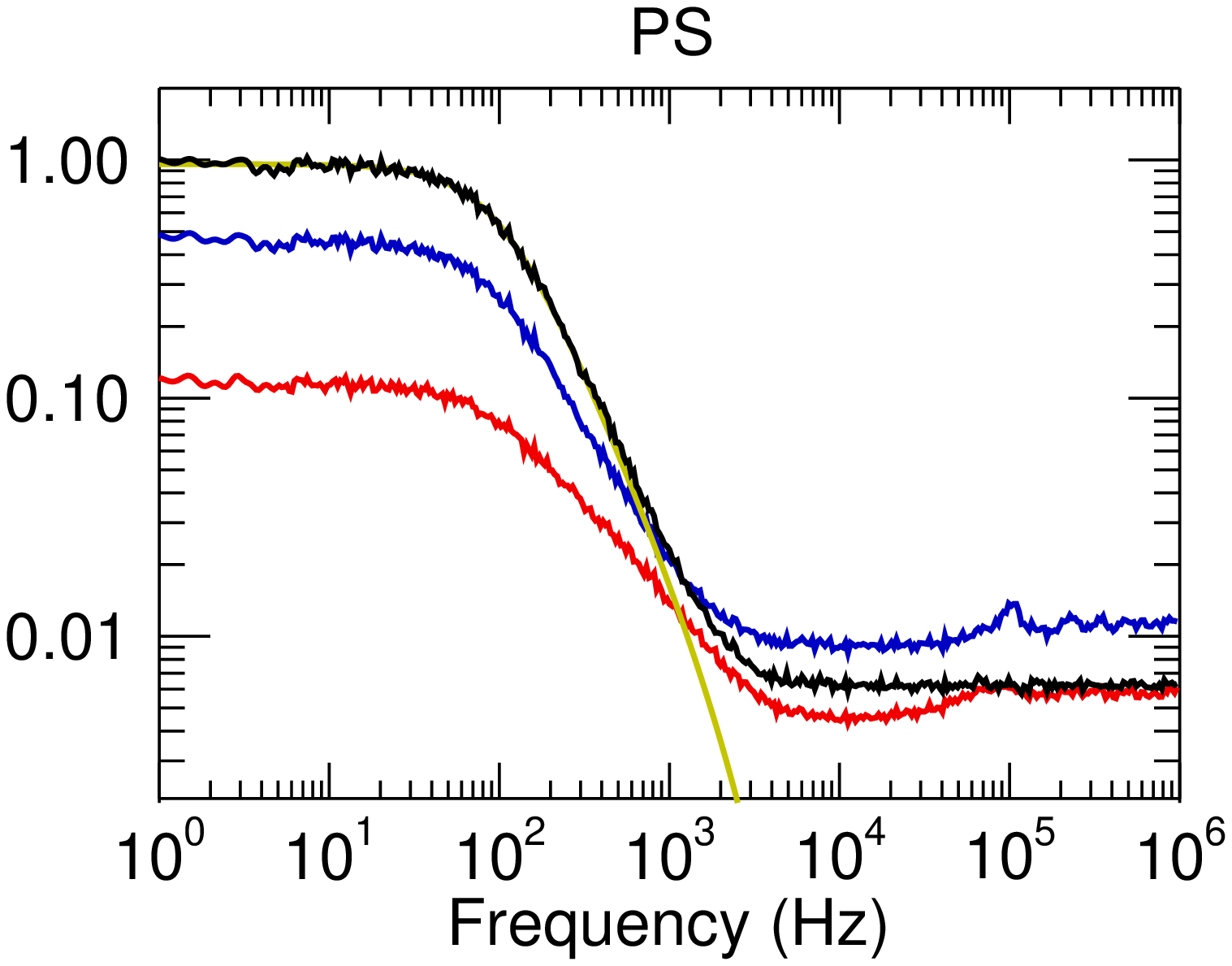}}
\end{minipage}
\begin{minipage}{0.5\linewidth}
  \center{\includegraphics[width=1.0\textwidth]{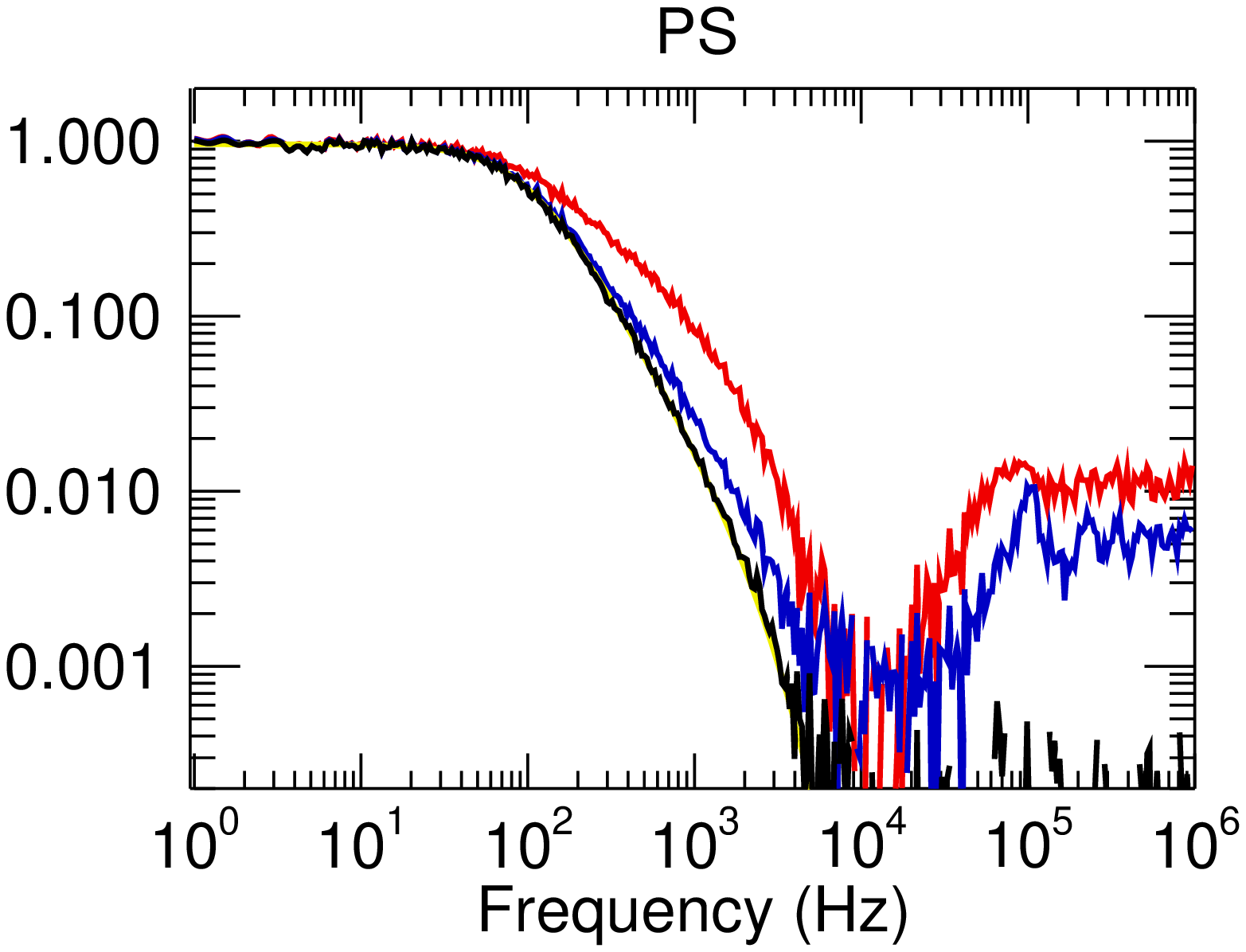}}
\end{minipage}
\caption{Power spectral bias due to fixed dead time effects. All spectra are averaged over 1000 blocks, each evaluated over a record length of 2\,s. Black: zero dead time, $\nu_0 = 71\,626\,Hz$, blue: non-paralyzable detector $\Delta t_d = 8\,\mu$s, $\nu_0 = 25\,732\,Hz$, red: paralyzable detector $\Delta t_d = 8\,\mu$s, $\nu_0 = 12\,986\,Hz$. Left: The different spectra have been normalized by the same factor that makes the von K\'{a}rm\'{a}n reference spectrum approach the value 1 at low frequencies. Right: The same spectra as in the left figure, all normalized individually to approach the value 1 at low frequencies.} \label{fig:Fig11}
\end{figure*}

It is important that the model for the LDA data includes all of the effects important to it, especially the residence times and a realistic dead time model. We have applied the two types of fixed dead time, the non-paralyzable and the paralyzable detector, to the computer generated randomly sampled velocity signal described above. The dead time model in Section~\ref{sec:2} refers to the non-paralyzable processor. Figure~\ref{fig:Fig11} shows that the spectral dead time bias can be significant even in the low frequency range if the dead times are sufficiently large (see~\cite{10} for a discussion) and also that the two types of dead time have very different effects on the measured spectrum. The left figure shows the different spectra normalized by the same factor that makes the von K\'{a}rm\'{a}n reference spectrum approach the value 1 at low frequencies. In the right figure all spectra have been normalized individually in the same fashion, for the sake of comparison. The yellow curve shows the von K\'{a}rm\'{a}n reference spectrum. The black curve shows the ideal case of zero dead time. The part of the spectrum above the constant offset represents the true spectrum, as can be seen from the collapse with the von K\'{a}rm\'{a}n spectrum. The blue curve shows the resulting spectrum in case of a non-paralyzable detector. The effect of the convolution with the $[\delta (f) - 2\Delta t_d \, \mathrm{sinc} (2 \pi f \Delta t_d)]$ function from Eq.~(\ref{eqn:specdead}) is clearly visible. The paralyzable detector (red curve) loses a lot of data due to the increased dead time, and the spectrum shows a loss of power, especially at low frequencies.

\subsection{Particle interference and residence time distribution}\label{sec:dtmodel}

Reverting to the measured LDA signals, it remains to describe which dead time model applies to the LDA burst-processor. We realize that the dead time problem is identical to the case of two (or more) particles being present within the measurement volume at the same time. In the description of LDA signal- and data processing it is often assumed that only one particle is present in the measurement volume at any one time. However, when we want to measure high frequency turbulence spectra it is desirable for practical reasons to have a high sample rate in order to obtain a high dynamic range between the true spectrum and the spectral offset. But this is exactly the condition that can lead to more than one particle in the measurement volume and thus to dead time effects. We must therefore consider in more detail what occurs in an LDA at high sample rates, where this condition may be violated.

\begin{figure*}[!h]
  \center{\includegraphics[width=1.0\textwidth]{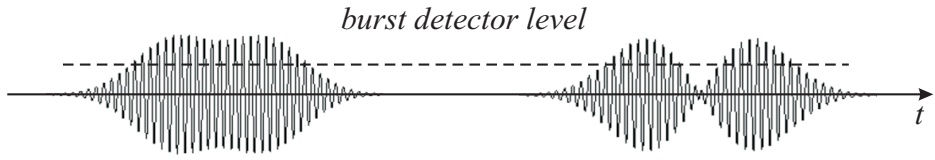}}
\caption{Case of two particles in measuring volume at the same time. Left: scattered Doppler bursts in phase, Right: scattered Doppler bursts out of phase.} \label{fig:Fig13}
\end{figure*}

Multiple particles in the measurement volume introduce phase-shifts in the Doppler signal that cannot be distinguished from the fluctuating velocity signal (c.f. Buchhave et al.~\cite{12}, George et al.~\cite{11}). In a burst detector, the interference between particles may result in longer or shorter bursts. Figure~\ref{fig:Fig13} shows a couple of situations for the high pass filtered Doppler signal: In the first one, two particles scatter Doppler modulated light bursts that happen to be in phase at the detector. The detector will see this as one sample with an extended residence time. This may be described as a case of a paralyzable detector. The second case is also two particles, both within the measurement volume, but scattering light that is out of phase at the detector. If the dip in the signal envelope due to the destructive interference is low enough, the system will see this as two particles arriving close to each other. Thus the result of a high particle arrival rate will be a broadening of the distribution of possible residence times and thus a wide range of possible dead times.

Fortunately, it will still be possible to describe the dead time effects since the residence times are measured for each sample, and the dead time distribution will therefore be known. However, this case cannot be described as a simple case of fixed paralyzable or non-paralyzable dead time. Instead we see the LDA case as an example of detection with a varying dead time, the measured residence time. When two particles interfere constructively as a sort of paralyzable dead time, the measurement is continued during the increased measurement time until the signal drops below the burst detector level. The dead time may even be increased by a fixed time delay during which the processor passes the data to the subsequent data processor and prepares for the next measurement.

These considerations form the basis for the dead time filter used to modify the computer generated velocity data. An algorithm checks if a particle arrives within the dead time of the previous particle and if so, extends the residence time with the residence time of the second particle. Note that the data point is not discarded; it comes in with a higher weight, but the small delay is lost. Note that the implementation of different common dead time filters, e.g., simply removing all measurements that overlap~\cite{BNT}, will differ from these measured dead times and are thus not representative of real signals. The application of this kind of filter is thus likely to overestimate dead time effects.

\subsection{Analytical description}\label{sec:analydescr}

\begin{figure*}[!h]
\begin{minipage}{0.5\linewidth}
  \center{\includegraphics[width=1.0\textwidth]{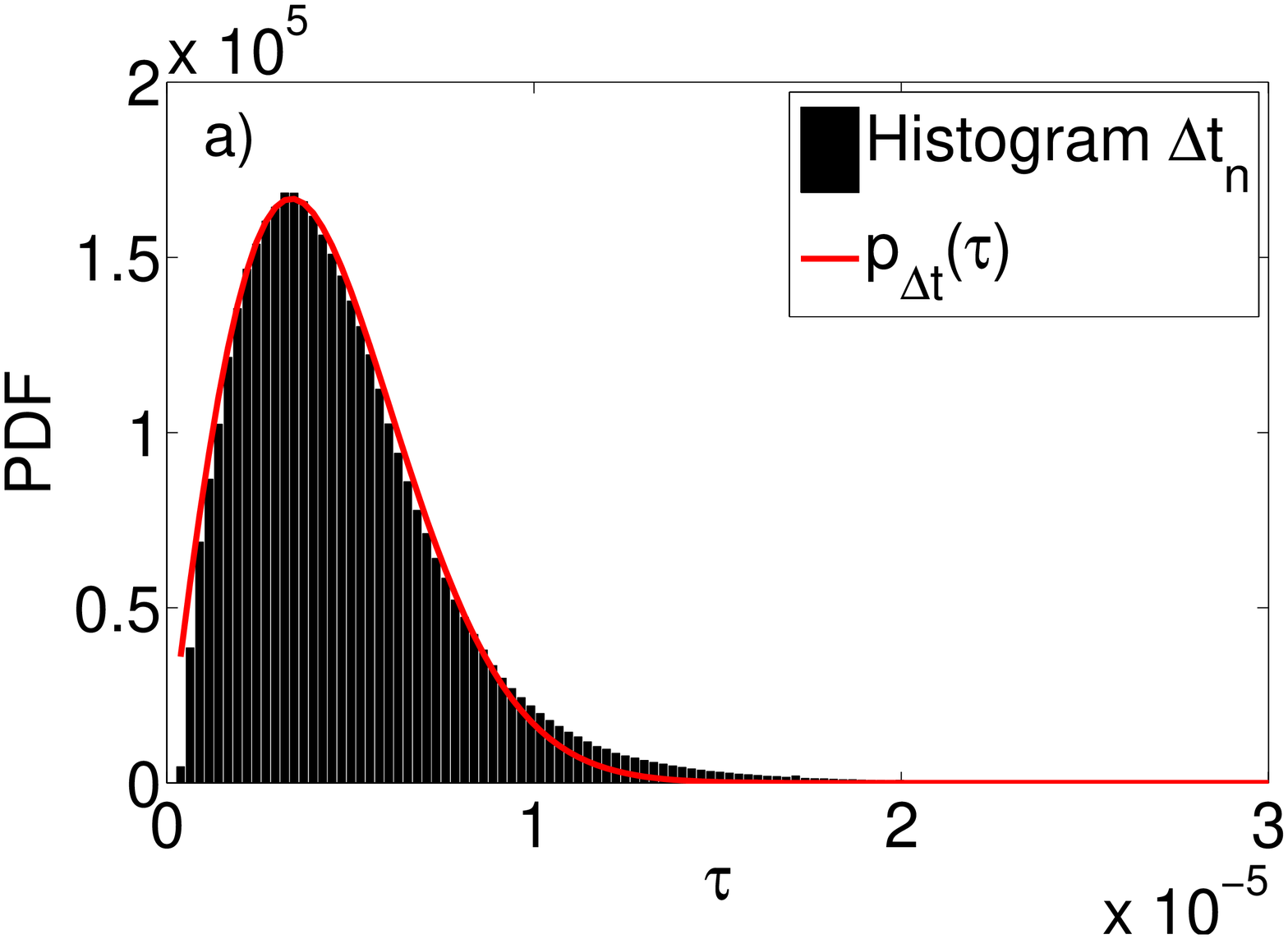}}
\end{minipage}\hspace{0.5cm}
\begin{minipage}{0.5\linewidth}
  \center{\includegraphics[width=1.0\textwidth]{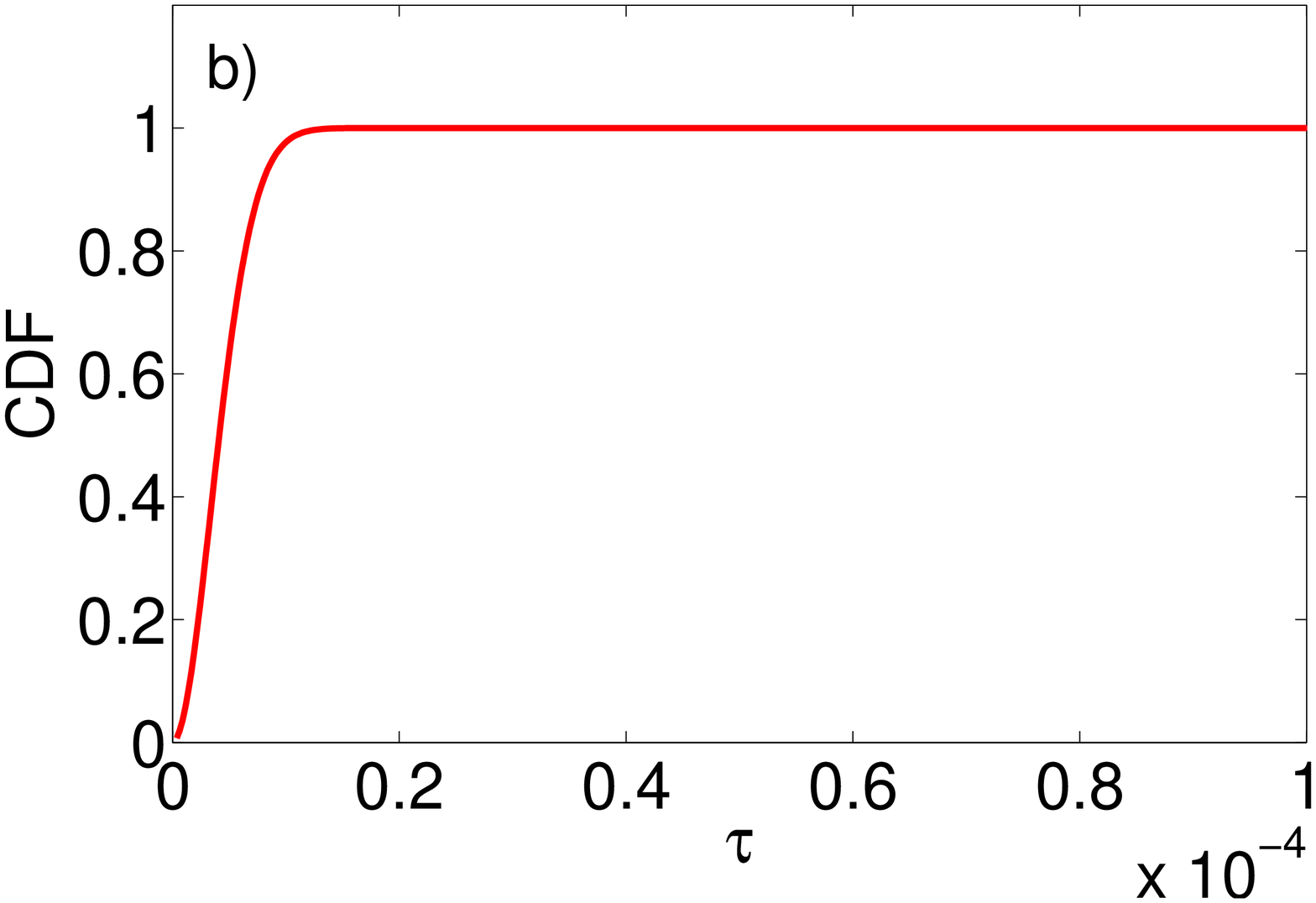}}
\end{minipage}
\caption{(a) Measured residence time probability density (black histogram) and matching Weibull density (red curve) and (b) cumulative Weibull distribution.} \label{fig:Fig14}
\end{figure*}

The probability density of the measured residence times depends on the flow properties, and we do not have an exact analytical expression. However, we do have the measured residence time data of our reference LDA power spectrum in Figure~\ref{fig:Fig12}, described more in detail in Sections~\ref{sec:method} and~\ref{sec:measuredspectrum}. The histogram in Figure~\ref{fig:Fig14}(a) shows the measured residence time probability density for our reference spectrum. It turns out that the so-called Weibull function (red curve in Figure~\ref{fig:Fig14}a) allows a nice fit to the measured residence time density with just two adjustable parameters:
\begin{equation}\nonumber
P(\Delta t_s) = \frac{k}{\lambda} \left ( \frac{\Delta t_s}{\lambda} \right )^{k-1} e^{-(\Delta t_s / \lambda)^k }
\end{equation}
The best fit to the measured data is $k = 1.875$ and $\lambda = 5.0\cdot 10^{-6}$.\footnote{Time between events is usually described by exponential distributions. However, the dead times may alter the shape of the distribution. The Weibull distribution was employed in this particular case since it provided a sensible fit to the measured residence time distribution. Note that the LDA dead time model presented in the present work does not depend on the type of distribution, but can be used with any distribution providing a good fit to measured data.}

The Weibull density is easily integrated to provide the cumulative Weibull distribution function, Figure~\ref{fig:Fig14}(b):	
\begin{equation}
C(\Delta t_s) = 1 - e^{-(\Delta t_s / \lambda)^k }
\end{equation}
which indicates how the dead time impacts on the ACF around the origin. We see no special significance of the Weibull distribution beyond the fact that it creates a nice fit in the current situation. Other flow experiments might require different methods, e.g., a polynomial fit.

\begin{figure*}[!h]
  \center{\includegraphics[width=0.5\textwidth]{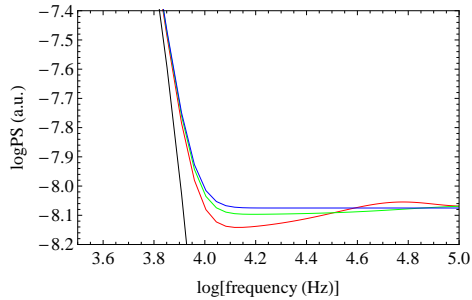}}
  \caption{von K\'{a}rm\'{a}n spectrum convolved with the weighted dead time response. Black: von K\'{a}rm\'{a}n reference spectrum, Blue: No dead time, Green: Weibull dead time distribution based on measured residence times, Red: Weibull dead time distribution based on measured residence times plus fixed dead time which is small compared to the residence times.} \label{fig:Fig15}
\end{figure*}

In practice the dead time, $\Delta t_d$, and the residence time, $\Delta t_s$, are nearly equal so we may consider the Weibull distribution to be a probability distribution describing the probability that a subsequent particle will arrive after the dead time and be registered as an independent measurement. Since we are dealing with (time) averaged statistics, we can use the Weibull density as an expression for a weighted distribution of the non-paralyzable dead time response given in Eq.~(\ref{eqn:specdead}) above. Then the resulting spectrum is simply:
\begin{equation}
S_{0,\,\Delta t_d,\, random}(f) = \int_{-\infty}^{\infty}S_{0,\Delta t_s}(f)P(\Delta t_s)\,d\Delta t_s
\end{equation}
The primary effect is that the integral smears out the original dead time window. To investigate the validity of this model, we have performed the convolution of the weighted dead time response with the von K\'{a}rm\'{a}n spectrum~(\ref{eqn:vkspectrum}) and arrived at the result shown in Figure~\ref{fig:Fig15}.

As can be seen, some of the features are explained in this derivation, but the Weibull density does not in itself sufficiently account for the dip in the spectrum. Thus the measured residence time density alone does not explain the dip. It appears that the processor does indeed have a finite data transfer time and thus the dead time is not exactly equal to the residence time in the present case. However, a small amount of fixed dead time added to the Weibull distribution does show the expected dip as can be seen in Figure~\ref{fig:Fig15}.

\section{Experimental setup}\label{sec:method}

The LDA consisted of Dantec two-component FiberFlow optics mounted on a two-axis traverse, two BSA Enhanced processors, and BSA Flow Software version 2.12. A $5\,W$ Coherent Innova 90 Argon-ion laser was used, running at $2\,W$. Only the $514.5\, nm$ wavelength was used, measuring the velocity component in the main flow direction. The system was operating in backscatter mode.

LDA measurements were performed in the streamwise direction in a turbulent axisymmetric jet at 30 jet exit diameters downstream of the jet nozzle. The jet exit velocity was nominally $30\, ms^{-1}$. The LDA system was placed to the side of the jet to minimize obstructions on the flow. The optical head was placed on a 3-axis traversing system and connected to the laser through fiber optics. The front lens had a focal length of 310\,mm, yielding an angle between the beams of 12.6$^{\circ}$, and the beam expander had an expansion ratio of 1.98. The beam width before the beam expander was 1.35\,mm and the wavelength of the employed component was 514.5\,nm. From optical considerations the measurement volume was deduced to have a maximum length of about 700\,$\mu$m and a maximum thickness of about 75\,$\mu$m. The overlap of the crossing beams was checked by placing a convex lens just ahead of the position of the measuring volume, projecting the magnified beams on a screen. The laser beams were then adjusted to maximize the overlap.

The main construction of the jet generator consisted of a cubic box of dimensions $58\times 58.5\times 59\, cm^3$, where the interior was stacked with foam baffles in order to damp out disturbances from the fan that supplied the generator with pressurized air. The box was fitted with an axisymmetric plexiglass nozzle, tooled into a 5th order polynomial contraction from an interior diameter of 6 cm to an exit diameter of $d=1\, cm$. The air intake was located inside the jet enclosure. For further details on the generator box, see~\cite{Gamard2,Jung}. The flow generating box rested on a rigid aluminum frame. The exit velocity was monitored via a pressure tap in the nozzle positioned upstream of the contraction and connected to a digital manometer by a silicon tube. The ambient pressure was monitored by an independent barometer. The enclosure utilized in the experiment was a large tent of dimension $2.5\times 3\times 10\,m^3$. The jet was positioned at the back of the enclosure. Under these conditions, the jet flow generated in the facility should be expected to correspond to a free jet up until $x/D = 70$.

From previous studies~\cite{4} the flow at the jet center line 30.3 jet exit diameters downstream of the jet exit has been estimated to have an integral scale of 0.011\,m, a Taylor microscale of $2.2\cdot 10^{-3}\,$m and the Kolmogorov scales can be estimated to $53\cdot 10^{-6}\,$m, respectively.

\section{Measured spectrum}\label{sec:measuredspectrum}

\begin{figure*}[!h]
\begin{minipage}{0.5\linewidth}
  \center{\includegraphics[width=1.0\textwidth]{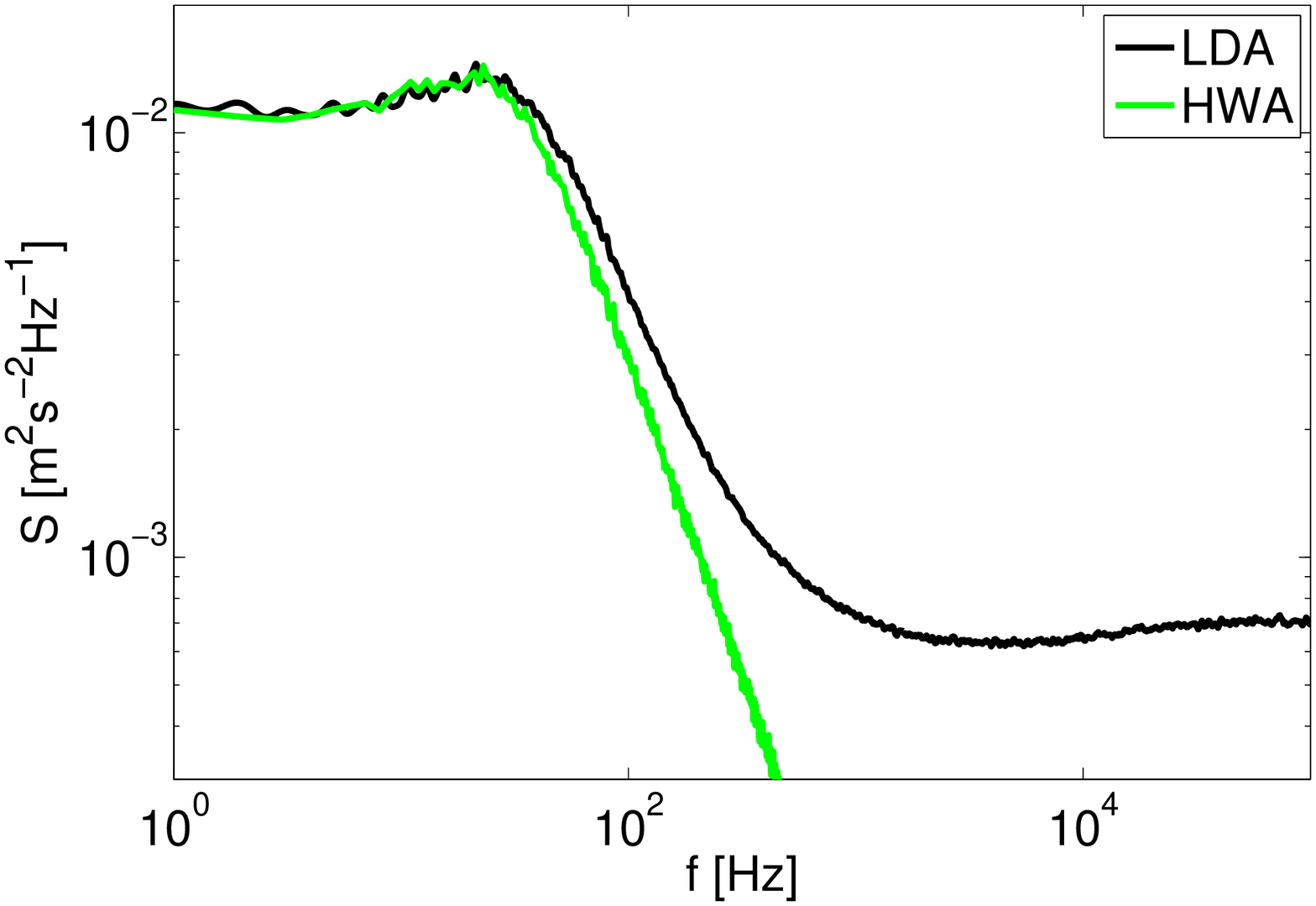}}
\end{minipage}\hspace{0.5cm}
\begin{minipage}{0.5\linewidth}
  \center{\includegraphics[width=1.0\textwidth]{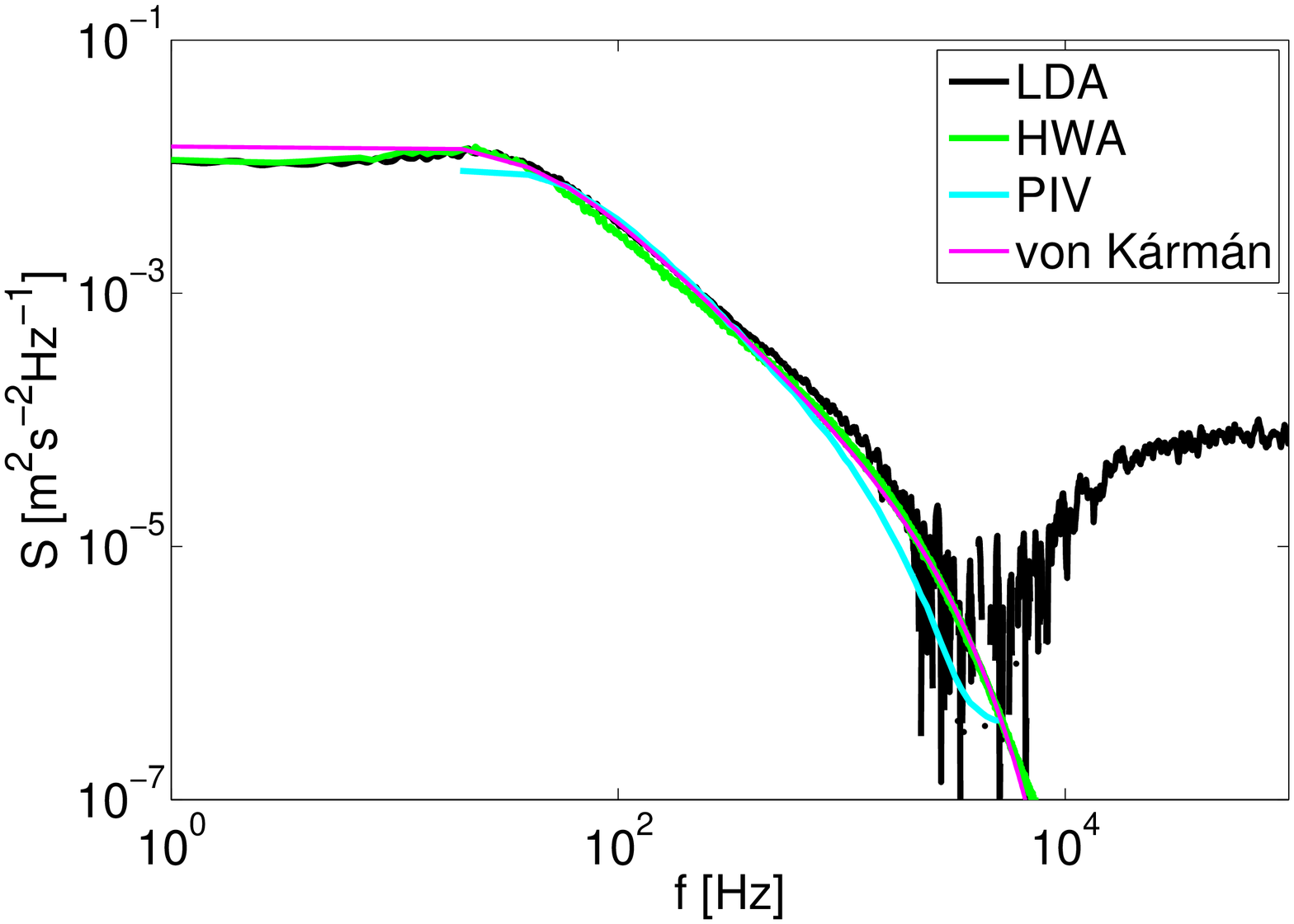}}
\end{minipage}
\caption{LDA measurement of turbulent power spectrum. (left) Comparison between LDA and HWA spectra and (right) also the corresponding spatial Stereoscopic Particle Image Velocimetry (SPIV) spectrum~\cite{Wanstrometal} and a von K\'{a}rm\'{a}n spectrum fitted to the HWA spectrum. From the LDA spectrum a constant offset has been subtracted.} \label{fig:Fig12}
\end{figure*}

The reference LDA power spectrum in Figure~\ref{fig:Fig12} (left) is measured at the jet centerline 10 jet diameters downstream with a resulting average data rate of $\nu = 9650\, Hz$. $10^7$ data points consisting of simultaneous values of arrival time, velocity component and residence time were used in the calculation. The LDA record was processed in $10^3$ blocks. The spectrum was computed using residence time weighting~(\ref{eqn:specest}) and comparison is made to a corresponding hot-wire power spectrum and a corresponding spatial Stereoscopic Particle Image Velocimetry (SPIV) spectrum~\cite{4,Wanstrometal}. Further, a von K\'{a}rm\'{a}n spectrum has been fitted to the spectrum obtained from the hot-wire measurements, see figure~\ref{fig:Fig12} (right) and Eq.~(\ref{eqn:vkspectrum}). Note that in the right figure a constant offset has been subtracted from the LDA spectrum to clarify that it does indeed collapse with the hot-wire spectrum up to around the measuring volume cut-off frequency. Further, the roll-off due to the larger measuring volume of the SPIV is visible at the highest frequencies.

We especially point out the following features of the LDA reference spectrum in the left Figure~\ref{fig:Fig12}:
\begin{itemize}
\item The spectral shape displays well the von K\'{a}rm\'{a}n like nature of the underlying turbulence structures.
\item The spectrum levels off at a relatively high constant asymptotic noise value.
\item The spectrum displays a dip where it breaks off to the constant level.
\end{itemize}

The constant noise level is rather high, which is characteristic of LDA measurements. As explained in Section~\ref{eqn:specdead}, the constant offset is a feature of randomly sampled power spectra and the level is dependent on the signal variance and the average data rate, see the first constant term in equation~(\ref{eqn:specdead}). One can remove the effect of this term by subtracting the self-products in Eq.~(\ref{eqn:specest})~\cite{10,4}. However, other factors may contribute to the constant offset: Noise due to phase fluctuations of the received light during the measurement, detector shot noise, electronic noise in detector and preamplifier and variations in the detected residence time due to variations in particle trajectories through the measurement volume. In Section~\ref{sec:CGdata} we have established separately that this contribution amounts to about $3\,dB$ increase in the noise floor.

The dip at the point where the spectrum levels off is indicative of a dead time effect. In the following Section we compare the reference turbulent power spectrum measured by LDA to computer generated data using the von K\'{a}rm\'{a}n model~(\ref{eqn:vkspectrum}) for the spectrum and the model for the dead time effect in an LDA described in Sections~\ref{sec:dtmodel} and~\ref{sec:analydescr}.

\section{Comparison to measurement}\label{sec:commeas}

\begin{figure*}[!h]
\begin{minipage}{0.5\linewidth}
  \center{\includegraphics[width=1.15\textwidth]{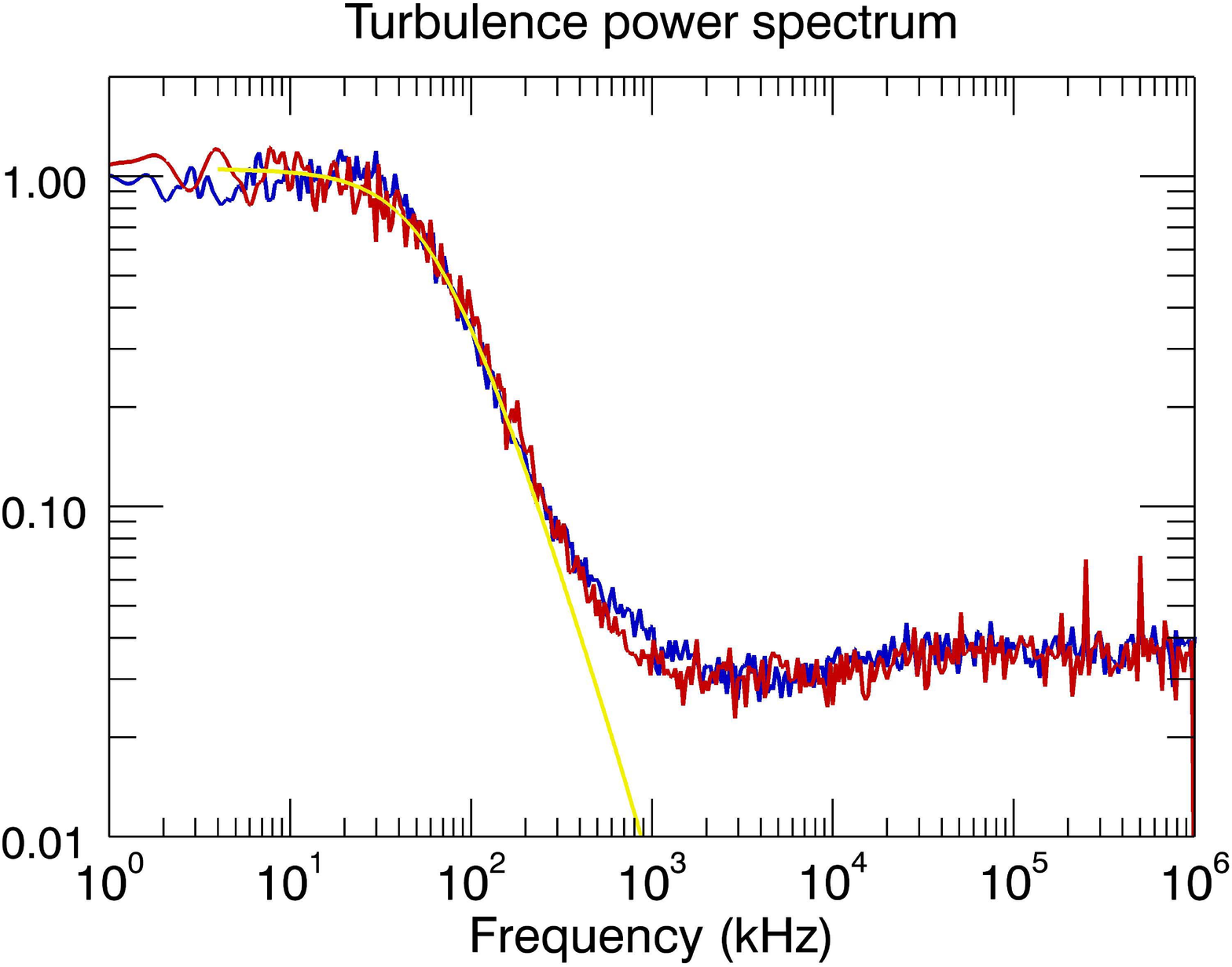}}
\end{minipage}\hspace{-0.45cm}
\begin{minipage}{0.5\linewidth}
  \center{\includegraphics[width=1.15\textwidth]{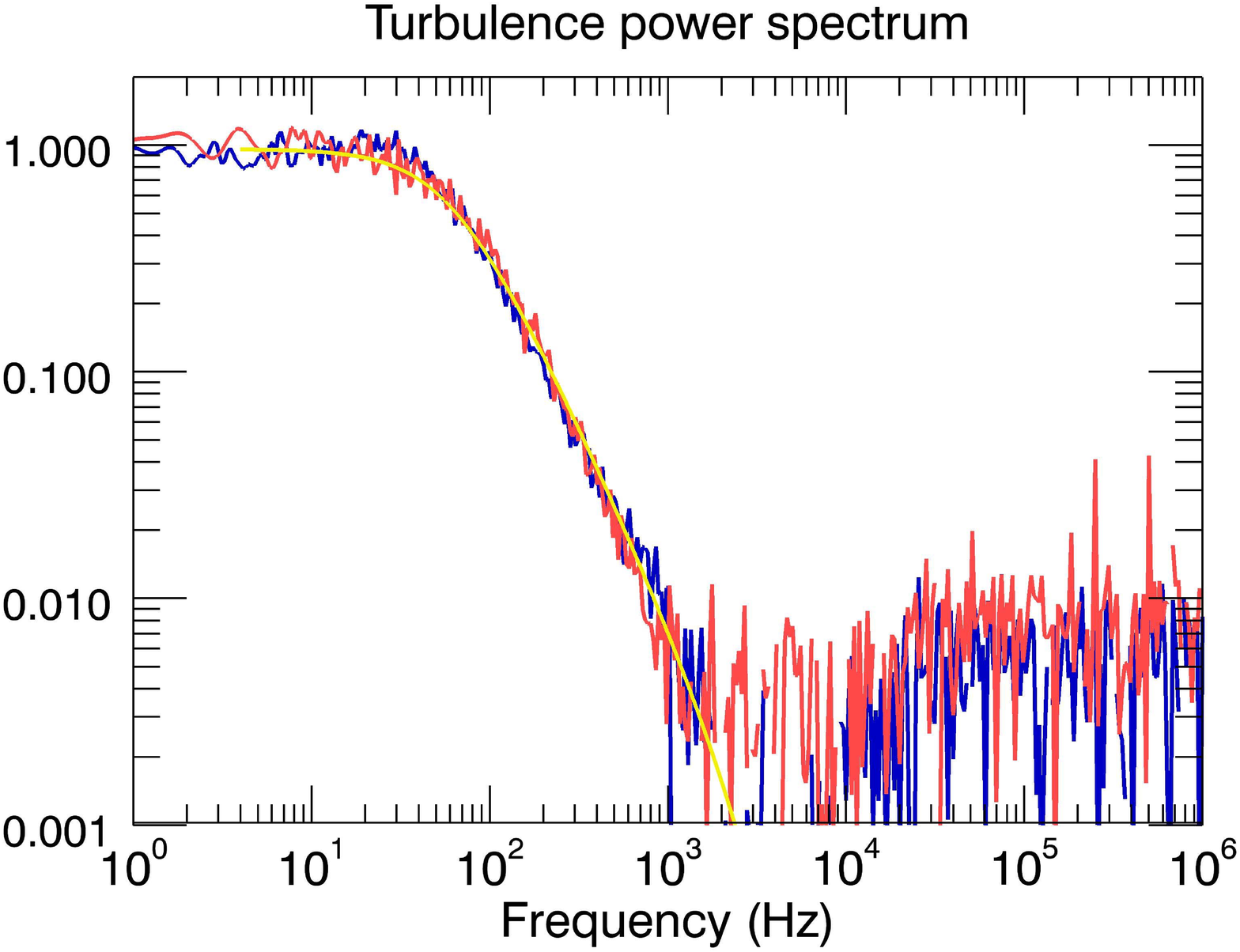}}
\end{minipage}
\caption{The measured turbulence spectrum (blue) and the CG spectrum with the measured Weibull residence time distribution plus a small fixed dead time (red). The best fit von K\'{a}rm\'{a}n spectrum of eqn. (\ref{eqn:vkspectrum}) is shown for reference (yellow). Left: The original measured and computer generated von K\'{a}rm\'{a}n spectra computed according to Eq.~(\ref{eqn:specest}). Right: The constant offset has been reduced to better illustrate the effects of the dead time on the power spectra.} \label{fig:Fig17}
\end{figure*}

We process the computer generated (CG) data and the measured velocity data through the same spectral estimator. As the real measurement volume diameter is a quantity that depends on a number of parameters such as particle size, detector/amplifier gain etc. we have adjusted the model measurement diameter $d_{MV}$ to give the best fit to the measured turbulence spectrum. The measurement volume diameter affects the width and location of the dip in the spectrum. Even with this adjustment, the offset level of the computer generated spectrum is lower than that of the measured spectrum, even with approximately the same data rate. We therefore add random white noise in the frequency domain before the frequencies are converted to a time series. Such noise may be detector shot noise, thermal noise in electronics or phase noise in the detected Doppler signal. Addition of this noise raises the constant noise floor. Finally, we add a small amount of fixed dead time ($4\,\mu s$) to the residence time distribution in Figure~\ref{fig:Fig14}. This additional dead time could be caused by a small finite processing or data transfer time added to the measured residence time. The two curves, the measured turbulence spectrum and the computer generated spectrum, now show excellent agreement, see Figure~\ref{fig:Fig17} LHS. In Figure~\ref{fig:Fig17} RHS we have subtracted a constant level to decrease the constant offset.

Thus, by adjusting the measurement volume size, data transmission time and ambient noise we have shown that our model may be used to describe the spectral bias introduced by the non-ideal properties of the LDA detector and signal processor.

\section{Summary and conclusion}

The effect of processor dead time on velocity power spectra computed by the direct method was analyzed, first as particular cases of either fixed non-paralyzable or fixed paralyzable dead times, acting directly on a model von K\'{a}rm\'{a}n spectrum. The analysis was made by means of a computer generated data series deduced from the same original von K\'{a}rm\'{a}n spectrum. The results confirmed the expected effects of the dead time: Reduced spectral power at low frequencies and an oscillation starting with a dip in the spectral power at the frequency range corresponding to the measurement volume cut off frequency. The sensitivity to the type of dead time was also shown to be significant for the tested case. The analysis was then extended to describe the laser Doppler anemometer, which introduces a number of additional challenges:
\begin{itemize}
\item The residence times, and hence also the dead times, are not constant, but have a wide distribution; this can, at least in the present case, be fitted to a Weibull distribution.
\item The burst processor appears to require a finite time for data transfer in addition to the residence time.
\item Due to the way the burst detector is constructed, interpreting bursts that may have any combination of constructive and destructive interference is not straightforward, but requires a more sophisticated dead time model than the simple fixed dead time model.
\end{itemize}
Despite these issues, a relatively simple, but realistic model for the LDA sampling process was developed and implemented in a computer program that provides simulated LDA data. This model is described in Section~\ref{sec:dtmodels} (and more specifically in Sections~\ref{sec:dtmodel} and~\ref{sec:analydescr}). The simulated LDA data were processed in exactly the same way as the measured turbulence data, and after addition of some random white noise and a small amount of data transfer time to the simulated data, excellent agreement was obtained. We have identified some of the problems with practical randomly sampled data, in particular for the LDA and the associated burst processor and have shown by comparing measured and simulated data that our model convincingly describes the LDA burst processor.

It is clearly advisable, for the purpose of reducing dead time effects, to reduce the measurement volume size compared to the flow scales, since the dip in the power spectrum is directly related to the probe volume cut-off frequency. Also, a small measurement volume reduces the loss of data rate due to dead time effects and allows a high average data rate, which reduces the spectral offset thereby increasing the dynamic range for the measured spectrum. Further, it is advisable to use fast processing/data transfer to reduce the effect of additional fixed dead time, which contributes to the unwanted oscillation in the spectrum at high frequencies.

We have focused in this paper on the so-called direct spectral estimator as it is a commonly used method providing fast evaluation and operating without additional correction methods. It is likely that the effect of dead time will be different in different estimators. This should be the topic for future investigations.

The authors hope that this work will provide clues to future improvement in LDA processors and strategies to better cope with the problems connected to the measurement of turbulent power spectra.

\section*{Acknowledgements}
The authors gratefully acknowledge the support of the Reinholdt W. Jorck og Hustrus Fond, journal no. 13-J9-0026 and Fabriksejer, Civilingeni{\o}r Louis Dreyer Myhrwold og hustru Janne Myhrwolds Fond, journal no. 13-M7-0039.

\end{document}